\documentclass[conference]{IEEEtran}
\IEEEoverridecommandlockouts
\usepackage{cite}
\usepackage{amsmath,amssymb,amsfonts}
\usepackage{algorithmic}
\usepackage{graphicx}
\usepackage{textcomp}
\usepackage{xcolor}
\usepackage[left=1.62cm,right=1.62cm,top=1.9cm]{geometry}
\def\BibTeX{{\rm B\kern-.05em{\sc i\kern-.025em b}\kern-.08em
    T\kern-.1667em\lower.7ex\hbox{E}\kern-.125emX}}
\begin{document}

\title{Computer Vision Based Neurology Brain Activity Rejection Architecture and Implementation}

\author{
    \IEEEauthorblockN{Zag ElSayed, \textit{Member, IEEE}}
    \IEEEauthorblockA{\textit{School of Information Technology} \\
    \textit{University of Cincinnati}\\
    Cincinnati, Ohio, USA}
    \and
    \
    \IEEEauthorblockN{Nathan Suer, Grace Westerkamp, Jack Yanchen Liu, \\ Makoto Miyakoshi, Craig Erickson, \textnormal{and} Ernest Pedapati}
    \IEEEauthorblockA{\textit{Division of Child and Adolescent Psychiatry} \\
    \textit{Cincinnati Children’s Hospital Medical Center}\\
    Cincinnati, Ohio, USA}
}

\maketitle

\begin{abstract}

The electroencephalogram (EEG) is a valuable and widely applied tool for investigating brain disorders and behavioral changes. It offers a minimally restrictive and non-invasive method. However, challenges in using EEG for cognitive development studies include temporal resolution, signal source localization, and EEG artifacts. Careful consideration of these factors is essential for informed application of EEG technology. Independent component analysis (ICA) effectively isolates source generator processes from signals recorded by multiple, adjacent EEG scalp electrodes. Although ICA decomposition requires manual inspection, selection, and interpretation of independent components (ICs), this process is time-consuming and demands expertise. Automated IC classification can achieve sufficient accuracy, expediting large-scale EEG research and enabling near-real-time applications in conjunction with brain activity rejection tasks, which are crucial for medical specialists. This study introduces an automated computer vision-based ICA rejection labeling tool compatible with widely used software interfaces like ICLabel and EEGLab. By automating the manual task, the proposed system reduces processing time by 7200-fold and achieves an accuracy of 89.45\%.

\end{abstract}


\section{Introduction}
\begin{figure}[htbp]
\centerline{\includegraphics[width =\linewidth]{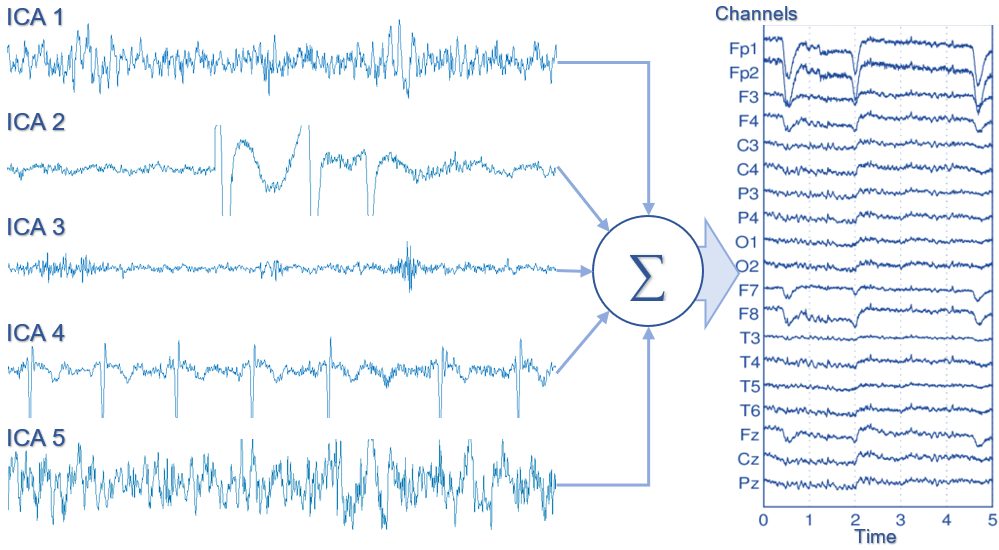}}
\caption{A portion of an EEG recording of ISO (10-20) with (5) dominant ICA Components.}
\label{fig1}
\end{figure}

Electroencephalography (EEG) remains a fundamental tool for studying brain activity, capturing the electrical signals generated by neuronal populations~\cite{khoshnevis2019applications}. Independent Component Analysis (ICA) has proven effective in separating mixed signals into independent components, revealing distinct neural sources. In this paper, we present a novel technique that leverages graphical representations derived from EEG-ICA to label specific brain activities, enhancing the interpretation ability and utility of EEG data, shown in Fig.~\ref{fig1}.

EEG monitors the electrical potential between two electrodes on the scalp, with evidence indicating the origin of this electrical signal ~\cite{pizzagalli2007handbook}. The EEG signal is context-dependent but spontaneous; the EEG produced during calm rest differs quantitatively from the EEG produced during cognitive functioning. The temporal resolution of the EEG signal is milliseconds. Postsynaptic alterations are instantly reflected in the EEG, which makes this technology exceptional for monitoring sudden changes in brain activity ~\cite{lukatch1996synaptic}. The robustness of electrical signals recorded at the scalp and the ease of use and non-invasive nature of the techniques used to obtain them make them valuable for research with younger people. On the other hand, getting high-quality signals usually takes a lot of training.

The literature currently has several techniques for removing EEG artifacts, and earlier research has focused chiefly on the manual or automatic identification of one or more different kinds of EEG artifacts. Independent Component Analysis (ICA) based techniques are frequently utilized in conjunction with other suggested ways to identify the artifacts~\cite{naik2011overview} effectively.

Various ICA algorithms have been created to estimate the distinct sources within a linearly mixed signal. However, different ICA algorithms have different estimating criteria, which could lead to different outcomes. Furthermore, the ICA algorithms do not name the estimated sources. Consequently, to clean the EEG signal and recover the ERP information, a criterion for choosing the appropriate ICA algorithm to achieve source separation and a technique for labeling various artifacts are needed.

\begin{figure}[htbp]
\centerline{\includegraphics[width =\linewidth]{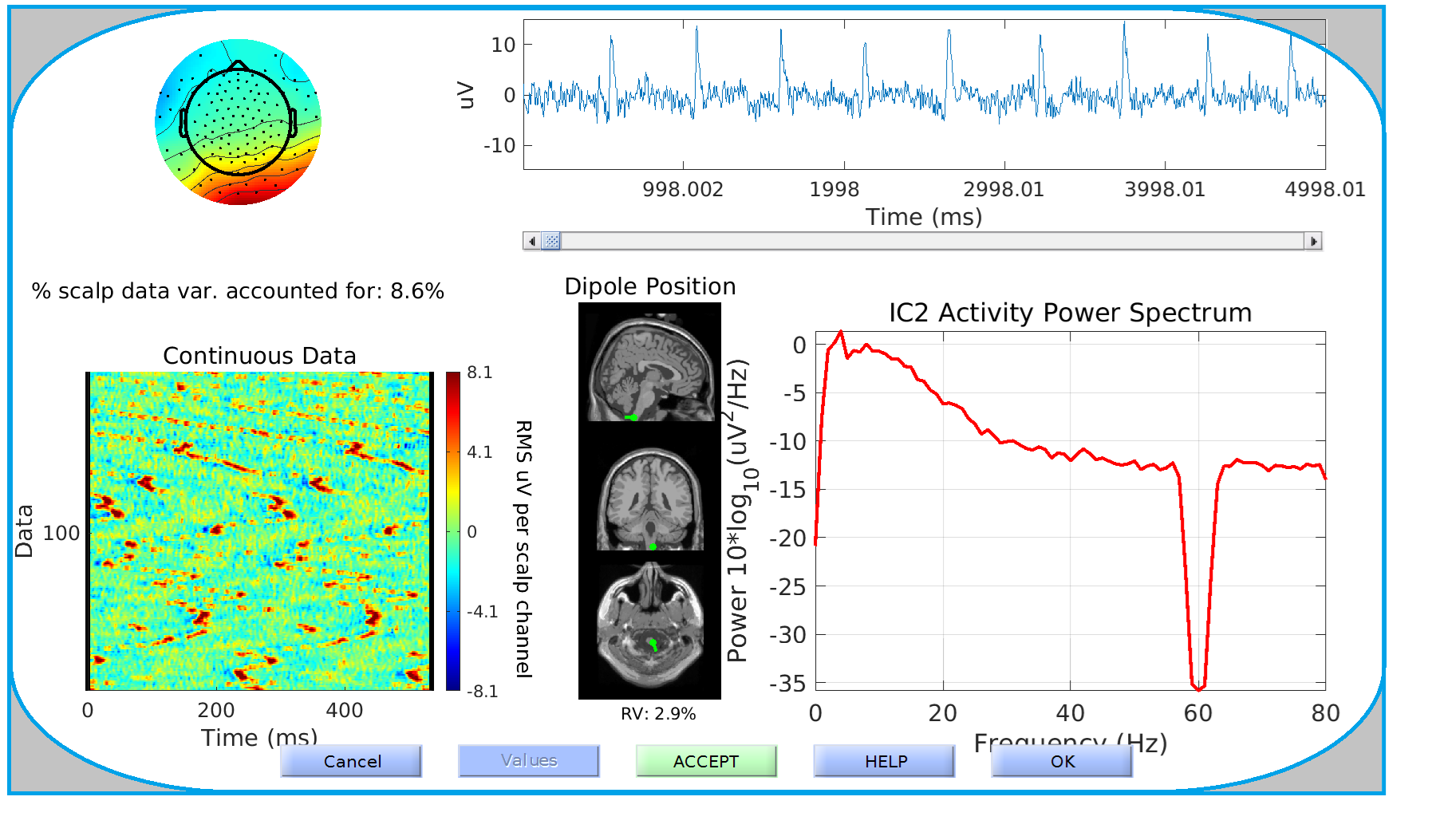}}
\caption{A sample screenshot of the ICA EEG Labeling Interface used by neurologists.}
\label{fig2}
\end{figure}

\begin{figure}[htbp]
\centerline{\includegraphics[width =\linewidth]{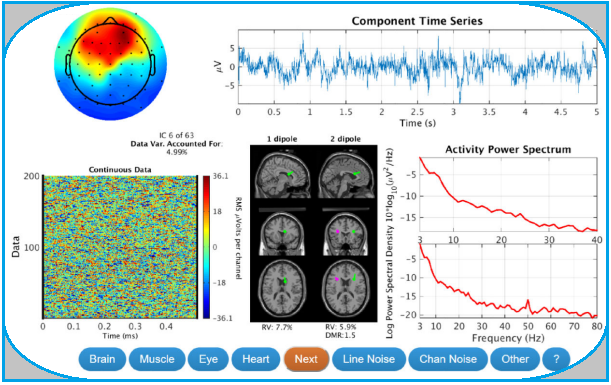}}
\caption{An IC labeling example from the ICLabel website~\cite{pion2019iclabel}.}
\label{fig3}
\end{figure}

\begin{figure}[htbp]
\centerline{\includegraphics[width =\linewidth]{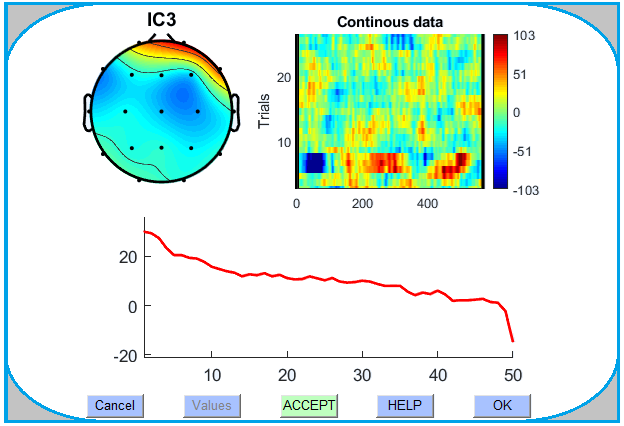}}
\caption{A sample screenshot of an IC labeling Matlab based tool~\cite{inproceedings}.}
\label{fig4}
\end{figure}

To obtain useful information from complicated neural signals and progress both basic and clinical neuroscience research, IC labeling of EEG recordings is essential for better knowledge of how the brain functions. IC labeling of EEG recordings holds significant importance in nonscientific research and clinical applications, especially for source separation, spatial localization, artifact removal, and cognitive function mapping. The mean labor time for a neurologist to perform such a talk using Matlab toolkit ($30 \pm10$ minutes) the large diversity is due to the nature of the visual analysis process and the human factor. An example of the ICA result interface that is being used by experts is shown in Fig.~\ref{fig2}, Fig.~\ref{fig3}, and Fig.~\ref{fig4}.

This work is the detailed extension of our previous work ~\cite{elsayed2024cnn}, and here we proposed a novel automated convolution neural network (CNN) model and tool for brain activity labeling using visual representation output for the most popular and currently applied Matlab tool (such as ICLabel) that neurologists and neuron scientists can use to expedite the analyses and save labor time and efforts for medical personnel. The proposed system was trained, tested, and verified using CCHMC Data sets where all EEGs were blinded and coded regarding participant, diagnosis group, and collection date. 

\section{Background}

\subsection{EEG and Brain Waves}

When conducting primary research on brain electrical activity during cognitive or effective processing, the EEG that is being presented here is commonly referred to as "quantitative EEG." Multiple sine waves cycling at various frequencies make up the EEG signal that was captured from the scalp. By breaking down the EEG into these distinct sine waves, Fourier analysis calculates the spectral power (measured in mean square microvolts) at each frequency. This yields information on the relative contributions of each frequency to the whole EEG spectrum at a given electrode site. It is believed that power represents the excitability of neural groupings

EEG is usually recorded during particular cognitive processing activities, and the recorded EEG is compared to the baseline or resting state EEG. One benefit of using EEG is that cognitive activities can last for a considerable amount of time—minutes or even seconds. For instance, many working memory task trials and the storage of stimuli for a subsequent recall or recognition memory task can yield EEG power and coherence values. EEG power and coherence can be averaged over trials if there are several of them. On the other hand, a single extended information processing session, like when encoding complicated stimuli, may be suitable.

Nevertheless, because the recorded signals in the raw EEG recordings are complicated and jumbled, originating from both neuronal and non-neural sources, they can be difficult to interpret. Because of the limitations of traditional EEG analysis methods in distinguishing various sources, sophisticated signal processing techniques have to be developed.
One increasingly effective method for handling the problems posed by mixed EEG signals is Independent Component Analysis (ICA). Using statistically independent components to represent individual neural or non-neural sources, ICA is a blind source separation approach that attempts to break down the observed signals. When various brain generators and artifacts are present in the recorded EEG data, this breakdown is especially helpful.

We are aware that there are specific difficulties in the Independent Component (IC) labeling procedure used in the analysis of Electroencephalogram (EEG) data, which can be summarized as the ambiguity in component interpretation, artifact contamination, cross-talk between neural groups, and spatial and temporal variability. However, the scope of this work is to provide practical and novel next-generation tools for brain activity using the current state-of-the-art.

\subsection{EEG Heatmap}
\begin{figure}[htbp]
\centerline{\includegraphics[width =\linewidth]{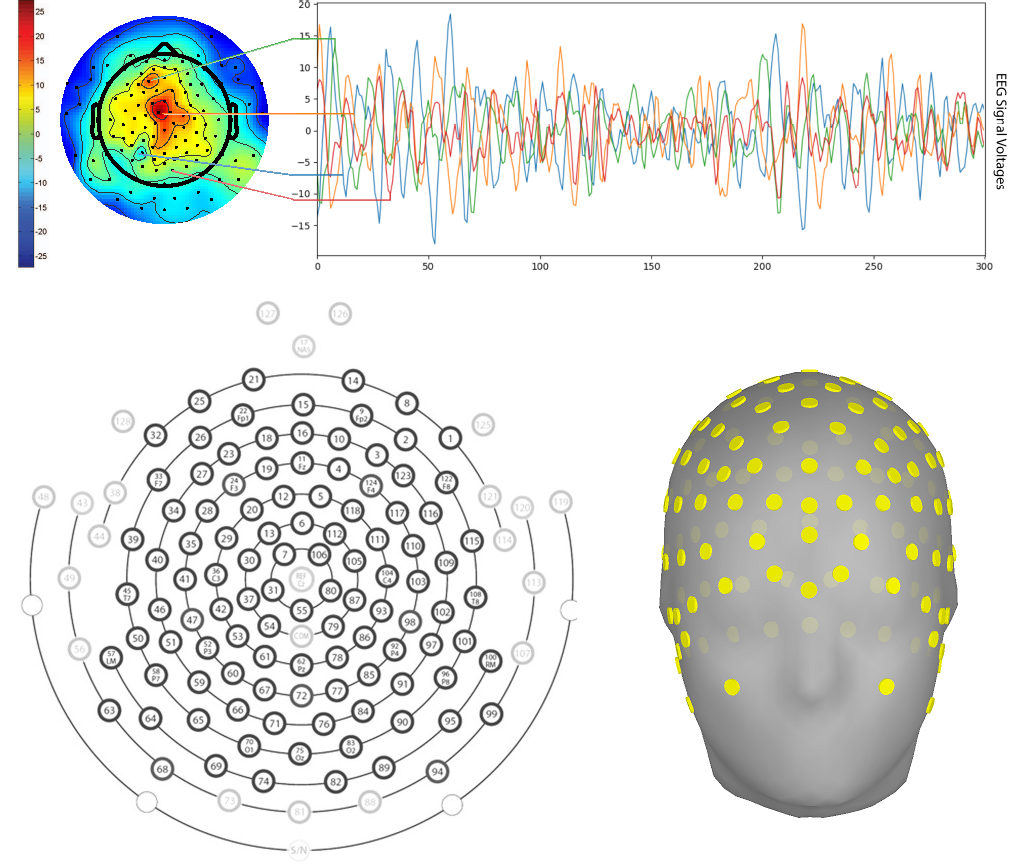}}
\caption{The EGI HydroCel Geodesic Sensor Net 128-Channel Map's layout for the total brain average of EEG biomarkers shows 108 chosen electrodes, with a heatmap showing the EEG nodes.}
\label{fig5}
\end{figure}

Brain activity can be inferred from EEG topographic maps. The utilization of brain mapping allows for the visualization of the brain's interconnection and functionality~\cite{7412206}. The determination of a functionally integrated relationship between geographically dispersed brain regions is aided by brain functional connectivity~\cite{LIU2023100070}~\cite{GYORFI2022135} and~\cite{s18124107}. The detailed components of the topographical representation is shown in Fig.~\ref{fig5}

\subsection{Machine Learning for EEG Data}
Incorporating machine learning methods into categorizing EEG heatmap images is a significant development in cognitive analysis and neuroscience research. Heatmap images created from EEG data visually depict brain activity at various temporal and spatial scales. However, advanced machine learning techniques must be applied for precise and effective categorization due to these images' high dimensional and inherent complexity.

Heatmap images created from EEG data are used to visualize neural activity spatially and discern brain response patterns and intensities. The dynamic interaction between different brain regions during cognitive tasks is commonly shown in these images, offering essential insights into the underlying neural processes~\cite{8713896}.
High-dimensional and multimodal elements that include details about brain activity's temporal and spatial aspects define EEG heatmap images. Machine learning approaches that can handle multidimensional inputs and capture complex patterns within and across modalities are necessary for the analysis of such complicated data~\cite{8972542} where multidimensional inputs machine learning top state-of-the-art techniques are listed as:
\begin{itemize}
    \item k-Nearest Neighbor in~\cite{al2018predicting}~\cite{antoniades2018deep}~\cite{gunay2018eeg}.
    \item Support Vector machine in~\cite{sai2017automated}~\cite{amin2017classification}~\cite{hosseini2020multimodal}.
    \item Linear and nonlinear Regression Models~\cite{roy2018deep}~\cite{kingsford2008decision}~\cite{rajaguru2017non}.
    \item Artificial Neural Networks~\cite{chiarelli2018deep}~\cite{hramov2017classifying}~\cite{lee2017cross}.
    \item Naive Bayes Application~\cite{sharma2018epileptic}~\cite{amin2017classification}~\cite{mumtaz2018machine}.
    \item Random Decision Forests~\cite{anastasiadou2017unsupervised}~\cite{bose2017eeg}~\cite{weichwald2014decoding}.
\end{itemize}

Additionally, feature extraction techniques are frequently used to extract pertinent data from the high dimensional input space when using machine learning algorithms on EEG heatmap images.

CNNs have demonstrated remarkable success in image classification tasks, making them well suited for EEG heatmap image analysis~\cite{jiang2021detecting}. The ability of CNNs to automatically learn hierarchical features from raw pixel values enables the extraction of spatial patterns and structures within the EEG images, facilitating accurate classification.
CNNs consist of layers designed to learn hierarchical features from input images automatically. Common layers include convolution layers, pooling layers, and fully connected layers.

Pre-existing models trained on extensive picture datasets, such as ImageNet, can be adjusted on smaller EEG datasets by fine-tuning. This enables the model to utilize the acquired features. Additionally, to address the frequently restricted availability of labeled EEG heatmap images, data augmentation techniques are employed to increase the size of the dataset~\cite{lashgari2020data} artificially. Randomly applying rotations, flips, and translations is a typical technique to introduce variances in the input data, which improves the model's capacity to generalize to unseen samples, which is critically beneficial for CCN-based model training.

Parameters such as learning rates, dropout rates, and filter sizes are tuned through grid or random search to find the optimal configuration for the EEG heatmap image classification task~\cite{poernomo2018biased}.
Choosing an appropriate loss function is essential for training the CNN. For labeled EEG heatmap images, categorical cross-entropy is commonly used for multi class classification tasks~\cite{pezzano2021cole}.

Interpretability in neuroscience is crucial for gaining insights into the neural processes underlying cognitive states. CNNs can be enhanced with interpretability techniques such as attention mechanisms or saliency maps, highlighting the regions in the EEG heatmap images that contribute most to the classification decision.

Challenges in applying CNNs to EEG heatmap image classification include the need for diverse and well-labeled datasets, addressing class imbalances, and ensuring generalizing of models across different experimental conditions. Future directions involve exploring novel architectures and advancing interpretability techniques to enhance the reliability of CNN-based classifications.

\section{Proposed Architecture Method}
\begin{figure}[htbp]
\centerline{\includegraphics[width =\linewidth]{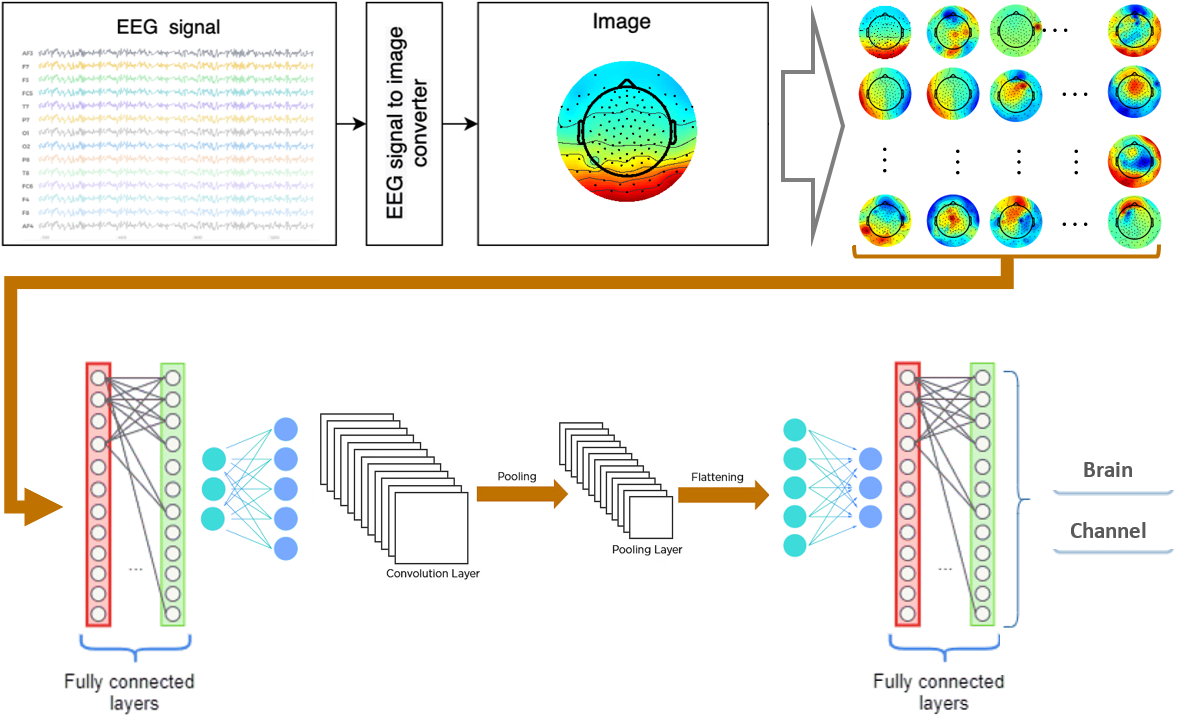}}
\caption{The proposed architecture process diagram}
\label{fig7}
\end{figure}

The proposed architecture is a novel method of automated rejection of the EEG ICA analysis of the medical personnel, and neurology scientists used to perform manually, and it then takes approximately 30 minutes per patient per recording. The proposed architecture is a system that can be tapped into an established ICA rejection pipeline process without disturbing or altering the clinical setup or the devices involved; the architecture is shown in Fig.~\ref{fig7} 
The architecture is based on a CNN-based model using the EEG topographic heatmap output report images.

Regrading input image generation, specifically for the color map selection, the power spectral density of each segment channel was calculated using the Welsh method, employing a Hamming window. The power within the six frequency bands was calculated as follows: 0.1–4 Hz (delta, $\delta$), 4–8 Hz (theta, $\theta$), 8–14 Hz (alpha, $\alpha$), 14–30 Hz (beta, $\beta$), 30–47 Hz (low gamma, $\gamma_1$), and 47–64 Hz (high gamma, $\gamma_2$)

\begin{figure}[htbp]
\centerline{\includegraphics[width =\linewidth]{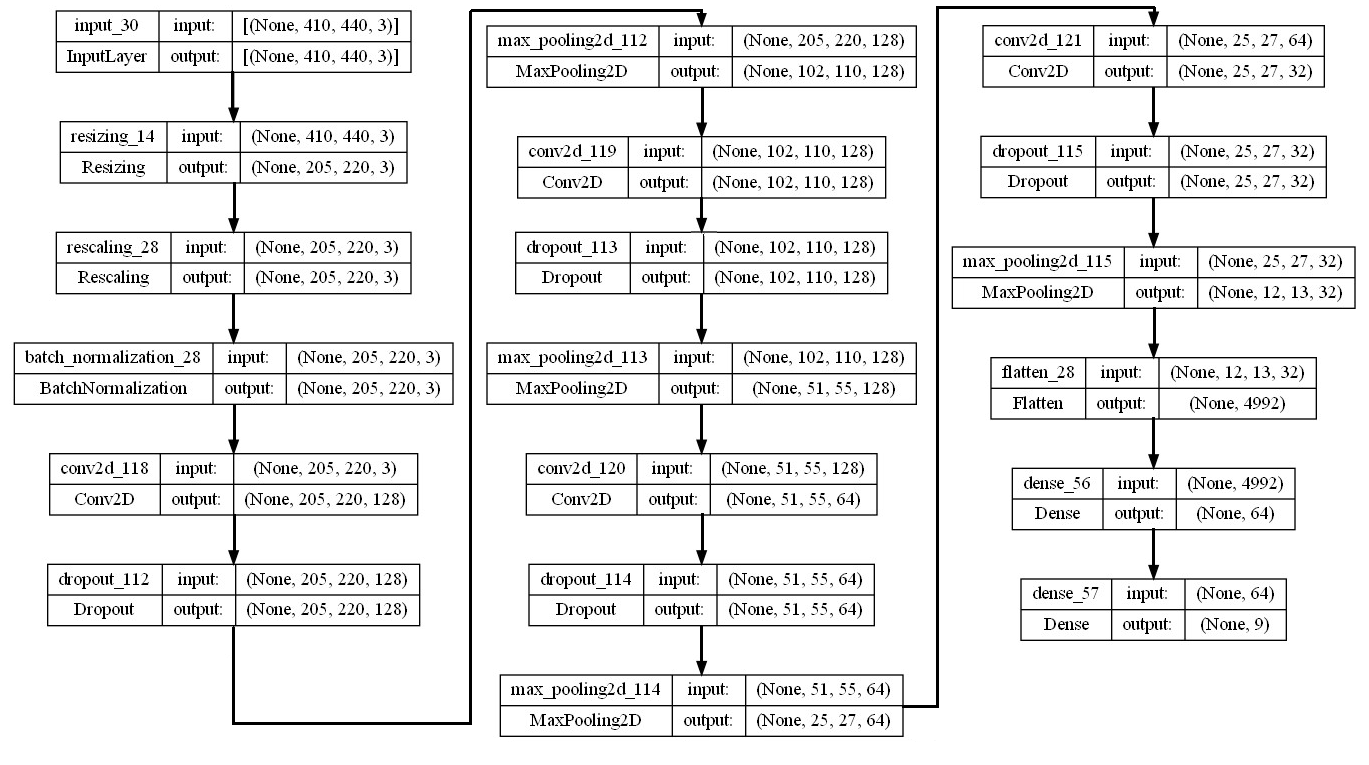}}
\caption{The detailed CNN Module Architecture diagram generated from Keras Library.}
\label{fig6}
\end{figure}

The architecture is based on a CNN-based model using the EEG topographic heatmap output report images. The power spectral density of each segment channel was calculated using the Welsh method, employing a Hamming window. The power within the six frequency bands was calculated as follows: 0.1–4 Hz (delta, $\delta$), 4–8 Hz (theta, $\theta$), 8–14 Hz (alpha, $\alpha$), 14–30 Hz (beta, $\beta$), 30–47 Hz (low gamma, $\gamma_1$), and 47–64 Hz (high gamma, $\gamma_2$).
The ultimate aim of this study is to reduce the complexity and power consumption of the proposed system for future hardware implementation prospectus. Thus, within the N.N design, we examined three activation functions: ReLU, Leaky ReLU, and Binary Step functions against the accuracy and power consumption. where the activation formulas are listed in~\eqref{eq1}~\eqref{eq2} and~\eqref{eq3}, respectively.
\begin{equation}
f(x) = max (0,x)\label{eq1}
\end{equation}

\begin{equation}
f(x)=max(0.01*x , x)\label{eq2}
\end{equation}

\begin{equation}
f(x)=
\begin{cases}
    x & \text{if $x < 0$}\\
    \alpha x & \text{if $x \geq 0$}    
\end{cases}
\label{eq3}
\end{equation}

where $x$ the input value to the neural calculation cell and $\alpha$ is the learning the parameter(booster).The result of optimizing the accuracy and power showed that the best performance was produced using the ReLU function. The CNN model is shown in  ~\ref{fig6}

\begin{figure}[htbp]
\centerline{\includegraphics[width =\linewidth]{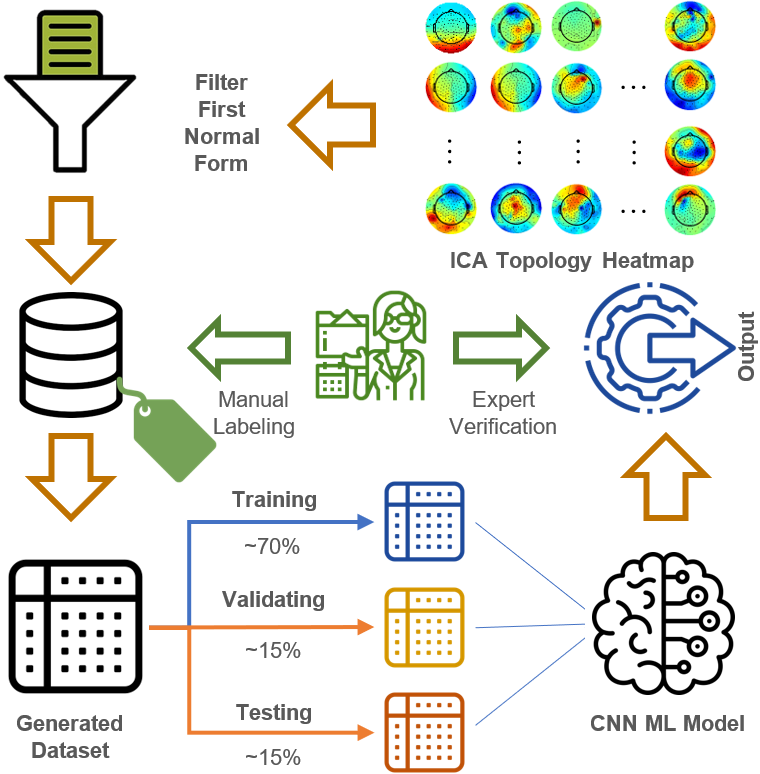}}
\caption{The data transformation process for the machine learning model and the architecture flow diagram.}
\label{fig8}
\end{figure}

For training and testing, clinical trials were performed and data collection was done at Cincinnati Children's Hospital Medical Center (CCHMC); the data was transformed into datasets for training, verifying, and testing the CNN model using a 128-channel HydroCel electrode net (Magstim EGI, Eugene, OR) and an EGI NetAmp 400 at a ${1000}$Hz sampling rate. Resulting in 5000 IC component interface dash images, extracted to heatmap $250\times250$ images. The rest of the process is illustrated in Fig.~\ref{fig8}

The clinical data was collected from nine (9) females and fifty (50) males; the age information and the diagnosis of the human subjects are listed in Table~\ref{tabel1} and Table~\ref{tabel2}, respectively.

\begin{table}[htbp]
\caption{Test subjects age statistcal data}
\begin{center}
\begin{tabular}{|c|c|c|c|c|c|}
\hline
\multicolumn{6}{|c|}{\textbf{Age}} \\
\cline{1-6}
\hline
\textbf{Min}&\textbf{1\textsuperscript{st} Que}&\textbf{Median}&\textbf{Mean}&\textbf{3\textsuperscript{rd} Que}&\textbf{Max}\\
\hline
6.25&11.96&14.58&19.90&25.58&44.58  \\
\hline
\end{tabular}
\label{tabel1}
\end{center}
\end{table}

\begin{table}[htbp]
\caption{Test subjects diagnosis}
\begin{center}
\begin{tabular}{|c|c|c|}
\hline
\textbf{Diagnosis}&\textbf{Count}&\textbf{Description} \\
\hline
TCD&17&Typically Developing Control  \\
\hline
FXS&19&Fragile X Syndrome  \\
\hline
ASD&22&Autism Spectrum Disorder  \\
\hline
DS&1&Down Syndrome  \\
\hline
\end{tabular}
\label{tabel2}
\end{center}
\end{table}

The ML model was constructed with four (4) convolution blocks with initial batch normalization ads scaling layers, padded with a flattening and two (2) dense layers at the end. Due to space limitations, we couldn't demonstrate the graphic representation, but It is available for public use on our GitHub page.

\section{Results Discussion and Analysis}
\begin{figure}[htbp]
\centerline{\includegraphics[width =\linewidth]{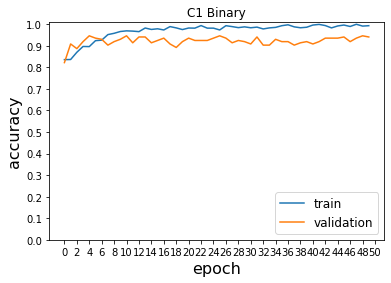}}
\caption{The CNN model training an validation accuracy graph for fifty epochs.}
\label{fig9}
\end{figure}

The proposed architecture was implemented and simulated using TensorFlow and Keras libraries on an Intel Core i9 CPU @ 2.40GHz, with NVIDIA GeForce RTX 2080S GPU. The model used 1,066,863 trainable parameters and six non-trainable parameters. The process runs over 50 epochs, and the accuracy of the training and validation graph is shown in Fig.\ref{fig9}.

During the testing phase, the lowest accuracy of perdition is 82.36\%, and the mean accuracy is 87.5\%. With an average time of execution, the ICs rejection task saved the neurologist time by $7200\times$. 

\subsection{Error Analysis}
This section presents the error analysis of the CNN model aimed for rejecting irrelevant brain activity. Understanding and analyzing these errors is crucial for improving the model’s performance and reliability.

We categorized the errors encountered by our model into the following types:
\begin{itemize}
    \item False Positives (FP): Instances where the model incorrectly classifies non-target heatmaps as target heatmaps.
    \item False Negatives (FN): Instances where the model fails to identify target heatmaps.
    \item True Positives (TP): Correctly identified target heatmaps.
    \item True Negatives (TN): Correctly identified non-target heatmaps.
\end{itemize}

\begin{figure}[htbp]
\centerline{\includegraphics[width =0.6\linewidth] {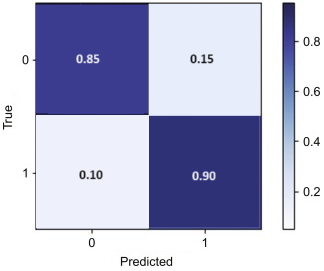}}
\caption{The confusion matrix for the proposed CNN model.}
\label{fig10}
\end{figure}

We evaluated the performance of our model using standard metrics such as Accuracy (the ratio of correctly predicted instances to the total instances), Precision (the ratio of true positive predictions to the total positive predictions), Sensitivity (The ratio of true positive predictions to the actual positives), F1 Score (the harmonic mean of precision and recall), and Confusion Matrix (A table summarizing the performance of the classification model). The confusion matrix for our CNN model is shown in Fig.\ref{fig10}.

Upon inspection, we observed that the model tends to confuse specific brain activity patterns, particularly in heatmaps with overlapping features. For instance, the model misclassified some non-target heatmaps with high-frequency noise as target heatmaps. This confusion can be attributed to the similarity in spatial patterns between noise and certain brain activities.

We conducted 50 experiments with 5000 inputs, addition to the automatic cross-validation with a manual expert inspection was added with every image to validate the results and provide a feedback path to the initial stages of the process as shown in Fig.\ref{fig8} regarding the Expert Labeling and Expert Validation tasks.
Based on the novelty of this and its sophisticate implementation we we only were able to compared it to the previous work~\cite{elsayed2024cnn}, the proposed CNN model demonstrated an 8\% improvement in precision and a 12\% reduction in false negatives. These improvements highlight the effectiveness of our model in accurately classifying ICA component heatmaps, especially with the enhanced feedback to the original dataset and the increased size of the labeled dataset of 150\% that we used in our second experiment.

\subsection{Limitations and Future Work}
We employed visualization techniques, such as saliency maps and heatmaps, to analyze the error patterns. These visualizations helped us identify specific regions in the heatmaps where the model's predictions were incorrect. Additionally, we performed statistical tests, confirming that the observed improvements were significant with a p-value less than 0.05. To address the identified errors, we implemented three strategies:
\begin{itemize}
    \item Data Augmentation, where we used techniques like rotation, scaling, and noise addition to improve model robustness and generalization.
    \item Architectural Changes, where we modified the CNN architecture by adding more convolution layers and using dropout to prevent overfitting.
    \item Hyperparameter Tuning, where we fine-tuned hyperparameters such as learning rate, batch size, and the number of epochs to optimize model performance.
\end{itemize}

The primary limitations of our error analysis include the size and diversity of the dataset, which may only partially represent the variability of all patients and brain activity. Future work will focus on expanding the dataset, exploring different model architectures, and integrating more advanced error mitigation techniques to enhance the model's performance further.

The authors are also aware of the bias of the hardware used and the quality of the image cropping. Additionally, we selected CNN architecture, which could appear too basic considering current advancements in computer vision. However, CNN was chosen for the simplicity of the machine learning model for the realistic further ASIC hardware implementation with limited memory size and clock for instruction set, as well as a proof of concept of the legibility of computer vision methods as a tool for diagnoses in the scientific domain of neurology where satiric linear statistical methods are the dominating and trusted methods.

\section{Conclusion}
This research explores the use of ML to classify IC-tagged EEG Images. The proposed method utilizes the spatial sensitivity of CNNs to extract features from intricate neural representations, offering an effective tool for identifying patterns linked to various cognitive states. Although this methodology shows potential, overcoming data limits and obstacles to interoperability is crucial to ensure its ongoing success. The effective incorporation of CNNs in the study of EEG not only improves our comprehension of regular brain activity but also has the possibility of being used in brain-computer interfaces and clinical diagnostics. For future work, we will implement the software package for this architecture for Matlab and Python as a library with a straightforward graphical user interface for medical personnel and neurologists. The journey ahead invites interdisciplinary collaboration and continuous innovation, and we remain optimistic about the profound impact that ML-based EEG IC analysis will have on shaping the future of neuroscience.

\bibliographystyle{ieeetr}
\bibliography{references}

\end{document}